\documentclass[a4paper,10pt]{article}

\usepackage[latin1]{inputenc}
\usepackage{amssymb}
\usepackage{epsfig}
\def\be{\begin{equation}}
\def\ee{\end{equation}}
\def\bea{\begin{eqnarray}}
\def\eea{\end{eqnarray}}

\begin{document}
\begin{center}

{\Large\bf Brane Gravitational Interactions \\ from 6D Supergravity}

\vspace{1cm}

{\large 
A. Salvio}\\

\vspace{0.6cm}

{\it {
IFAE, Universitat Aut\`{o}noma de Barcelona, \\08193 Bellaterra,
Barcelona, Spain \\  Email: salvio@ifae.es}} 

\end{center}

\begin{abstract}
We investigate  the massive graviton contributions to 4D gravity in a 6D brane world scenario, whose bulk field content can include that of 6D chiral gauged supergravity. We consider a general class of solutions having 3-branes, 4D Poincar\'e symmetry and axisymmetry in the internal space.  We show that these contributions, which we compute analytically, can be independent of the brane vacuum energy as a consequence of geometrical and topological properties of the above-mentioned codimension two brane world. These results support the idea that in such models the gravitational interactions may be decoupled from the brane vacuum energy.
\end{abstract}
\vspace{0.3cm}
{\small{\it Keywords}:\\ Field Theories in Higher Dimensions, Large Extra Dimensions, Supergravity models }

\section{Introduction}

Higher dimensional theories offer new avenues to address longstanding fine tuning problems. Regarding the gauge hierarchy problem, there are now other possible solutions in addition to 4D supersymmetry, such as the Randall-Sundrum \cite{Randall:1999ee,Randall:1999vf} and the Large Extra Dimensions scenario \cite{Arkani-Hamed:1998rs} (which requires, in its minimal formulation, at least two extra dimensions). Still the cosmological constant problem remains an unsolved issue in theoretical physics.

 A combination of the concepts of supersymmetry and (large) extra dimensions has been proposed \cite{Aghababaie:2003wz}  as a way to attack the cosmological constant problem, postulating two extra dimensions and  codimension two 3-branes. Whether this approach can be successful, it is still unclear (for criticisms and replies see e.g. \cite{Garriga:2004tq}-\cite{Burgess:2004dh}), however, an interesting property of this scenario would be  the prediction of the Kaluza-Klein (KK) scale at $10^{-3}$eV. One immediate consequence would be the onset of deviations from standard gravity at that scale, which corresponds to the submillimeter. Thus, a  natural question is whether the matching with experimental and observational tests of gravity can impose (additional) tuning of the brane tensions, once a 4D flat background solution is chosen.

The aim of the present paper is to answer this question by considering a realization of 6D supergravity called the Salam-Sezgin model\footnote{For an analysis of deviations from Newton's law in different types of 6D supergravity see \cite{Callin:2005wi}.} \cite{Salam:1984cj} (and its anomaly free \cite{seifanomaly} extensions \cite{Nishino:1984gk}). String theory derivations of this 6D supergravity have been provided \cite{Cvetic:2003xr} and its vacuum structure has been investigated in great detail. Gibbons, G\"uven and Pope (GGP) proved \cite{Gibbons:2003di} that the only smooth solution of the Salam-Sezgin model is the 4D flat unwarped space with internal spherical geometry and a non vanishing gauge field \cite{RandjbarDaemi:1982hi}. As soon as 3-branes are introduced, conical singularities are generated and the most general solution with 4D Poincar\'e symmetry and axial symmetry in the internal space has been derived \cite{Gibbons:2003di,Aghababaie:2003ar} (we consider this class of solutions here and refer to it as the GGP solution). There are no solutions of the Salam-Sezgin model having a curved 4D space with maximal symmetry and without singularities stronger than conical \cite{Burgess:2004dh,Tolley:2005nu}. These properties render the GGP solution an interesting
set of configurations, whose 4D spectrum has been indeed studied in a series of works \cite{Parameswaran:2006db,Burgess:2006ds,Parameswaran:2009bt}.

Here we focus on the minimal coupling between gravity and branes required by general covariance and compute the brane interactions mediated by massive KK gravitons. We derive their general form at the leading order in perturbation theory and their corrections to Newton's law on the brane. We show that these quantities can be independent of the brane tensions and we comment on the physical implications of this property. The origin of such independence is identified with the presence of codimension two branes on a compact space.

\section{The class of Models}

We consider a class of models in $D$ dimensions possessing 3-brane solutions with 4D Poincar\'e invariance. The complete (covariant) action is $S= S_B+S_b$, where $S_B$ is the bulk action (see below for specific forms) and $S_b$ is the brane one:
\be S_b=\int d^4 x \sqrt{-g}(-T+ I(x)), \label{Sb}\ee
where $g$ is the determinant of the brane metric, defined in terms of the higher dimensional metric $G_{MN}$ by \cite{Sundrum:1998sj}:
\be g_{\alpha \beta}\equiv G_{MN}(Y(x))\partial_{\alpha}Y^M(x) \partial_{\beta}Y^N(x). \label{branemetric}\ee
Here $M,N,...=0,...,D-1$ and $\alpha, \beta,...$ are the brane world volume indices: $x\equiv\{x^\alpha\}$. The functions $Y^M(x)$ give the position of the generic point $x$ on the brane in the higher dimensional coordinate system. The function $I(x)$ instead can be a generic functional of $g_{\alpha \beta}$ and additional brane fields $\chi_b$, but below we will consider it as a gravitational source localized on the brane. The constant $T$ is the brane tension that may include the vacuum energy produced by the brane fields. It is trivial to generalize this to an arbitrary number of branes: all we have to do is to add an index to the various quantities appearing in (\ref{Sb}), but from now on (unless otherwise stated) we suppress it. 

We are interested in the 6D case and in a class of models which can describe 6D Einstein-Yang-Mills systems (with a cosmological constant $\Lambda$), which we refer to as EYM$\Lambda$, and 6D N=1 gauged supergravities \cite{Nishino:1984gk,seifanomaly}. Also we will focus on the bosonic bulk dynamics, as, in our setup, only bosons couple to the brane sources. The bulk fields (which depend on all the space-time coordinates $X^M$) include, in addition to the higher dimensional metric, the gauge field $\mathcal{A}_M$ of a compact Lie group $\mathcal{G}$. In order to complete the bosonic part of 6D supergravity, one should also add a
dilaton $\phi$ and a 2-form field $B_{MN}$, which emerges from the
graviton multiplet and an antisymmetric tensor multiplet
\cite{Nishino:1984gk}. Moreover, 
concerning 6D supergravity, we shall assume that $\mathcal{G}$ is
a product of simple groups that include a $U(1)_R$ gauged R-symmetry. 
In general one can also add some hypermultiplets
\cite{Nishino:1984gk}, which turn out to be important to cancel gauge
and gravitational anomalies \cite{seifanomaly}. In the
bosonic sector this leads to additional scalar fields $\Phi^{\alpha}$; however, from
now on we consistently set $\Phi^{\alpha}=0$.

The bulk action is
\bea S_B=\int d^6X
\sqrt{-G}\left\{\frac{1}{\kappa^2}\left[R-\frac{1}{4}\left(\partial
      \phi\right)^2\right] 
-\frac{1}{4}e^{\phi/2}F^2 \right. \nonumber \\ \left.
-\frac{\kappa^2}{48}e^{\phi}H_{MNP} H^{MNP} -\mathcal{V}(\phi)\right\},
\label{SB} \eea
where $\kappa$ is the six-dimensional Planck scale, $F_{MN}$ is the usual gauge field strength and 
\be H_{MNP} \equiv \partial_M B_{NP} +
F_{MN}{\cal A}_P - \frac{g}{3}{\cal A}_M\left({\cal A}_N
\times {\cal A}_P \right) + \,\rm{2 \,\, cyclic \,\, perms} \,, \label{HKR}\ee
where $g$ is the gauge
coupling that  in fact represents a collection of
independent gauge couplings, which may include that of a $U(1)_R$
subgroup, $g_1$.
In the
supersymmetric model the dilaton potential is fixed to be $\mathcal{V}(\phi)=8\,g_1^2
\,e^{-\phi/2}/\kappa^4$.  With obvious truncations we can recover the EYM$\Lambda$ model: $\phi=0$, $ H_{MNP}=0$ and $\mathcal{V}(0)=\Lambda$.

The GGP solutions \cite{Gibbons:2003di} are
\bea ds^2&=&e^{A(u)}\left(\eta_{\mu \nu}dx^{\mu}dx^{\nu}+du^2\right)
+ e^{B(u)} \, \frac{r_0^2}{4} \, d\varphi^2 \, ,\nonumber \\
 e^A&=&e^{\phi/2}=\sqrt{\frac{f_1}{f_0}}, \quad e^B=4\,\alpha^2
e^A\frac{\cot^2(u/r_0)}{f_1^2},\nonumber\\
\mathcal{A}&=&-\frac{4\alpha}{q\kappa f_1}\, Q\,
d\varphi,\label{GGPsolution}\eea
where $r_0^2\equiv \kappa^2/(2g_1^2)$, $u$ is a compact coordinate ($0\leq u \leq \overline{u}\equiv \pi r_0/2$), $\varphi$ is an angular coordinate ($\varphi \sim \varphi +2\pi$), $\alpha$ and $q$ are real numbers and $Q$ is a generator
of a $U(1)$ subgroup of a simple factor of $\mathcal{G}$, satisfying
Tr$\left(Q^2\right)=1$. Also,
\be f_0\equiv 1+\cot^2\left(\frac{u}{r_0}\right), \quad f_1 \equiv
1+\frac{r_0^2}{r_1^2}\cot^2\left(\frac{u}{r_0}\right), \label{GGPsolution2}\ee
with  $r_1^2\equiv 8/q^2$. These solutions are supported by two branes at $u=0$ and $u=\overline{u}$, with $Y^{\mu}=\overline{Y}^{\mu}=x^{\mu}$. As $u \rightarrow 0$ or $u \rightarrow
\overline{u}$, the metric tends to that of a cone, with respective
deficit angles
\be \delta = 2\pi \left(1-|\alpha| \, \frac{r_1^2}{r_0^2}\right)
\quad \mbox{and}\quad
\overline{\delta}=2\pi\left(1-|\alpha| \right) \, . \label{deltadeltabar}\ee
 The two brane tensions $T$ and $\overline{T}$ are related to the deficit angles \cite{chenlutyponton}: 
$ T=2\delta/\kappa^2
$ and $\overline{T}=2\overline{\delta}/\kappa^2$.
Also, the internal space corresponding to the GGP has the $S^2$ topology (its Euler number
equals 2). Observe that the warp factor in (\ref{GGPsolution}) satisfies
\bea && e^A\stackrel{u\rightarrow 0}{\rightarrow}constant \neq 0,\quad
e^A\stackrel{u \rightarrow \overline{u}}{\rightarrow}
constant \neq 0, \label{Aproperty1}\\&& \partial_u e^A\stackrel{u\rightarrow 0}{\rightarrow} 0 \quad
\partial_u e^A\stackrel{u \rightarrow \overline{u}}{\rightarrow} 0. \label{Aproperty2}\eea
The relation between the 6D Planck scale $\kappa$ and our observed
4D Planck scale $\kappa_4$ is $ V_2/\kappa^2=1/\kappa_4^2$, where the volume $V_2$ is given by
$ V_2=\pi r_0\int du \,e^{(3A+B)/2}$.
For the GGP solutions we have $ V_2=4\pi \alpha \left(r_0/2\right)^2$.

We now discuss the independent parameters of our model (in the absence of sources). Before the compactification the action $S_B+S_b$, given  by (\ref{SB}) and (\ref{Sb}) with $I=0$, has the following independent parameters
\be \kappa, g', g_1, T, \overline{T}, \label{parametersBC} \ee
where $g'$ represent the collection of independent gauge couplings different from $g_1$. Solving {\it locally} the equations of motion (EOMs) with the GGP configuration does not constraint (\ref{parametersBC}). However, the fact that the internal space has a spherical topology imposes a {\it topological} constraint on the tensions. This is the usual Dirac quantization condition, which for a field interacting with  the background gauge field $\mathcal{A}$ through a unit charge gives 
\be \left(1-\frac{\kappa^2}{4\pi}T\right)\left(1-\frac{\kappa^2}{4\pi} \overline{T}\right)\left(\frac{\overline{g}}{g_1}\right)^2=N^2,\label{topological}\ee
where $N$ is an integer (the monopole number) and $\overline{g}$ is the gauge constant associated with $\mathcal{A}$. So after the compactification we have the same number of independent parameters, which can be taken as 
\be \kappa_4, r_0, g',T, \overline{T}, \label{parametersAC}\ee
but with the topological constraint (\ref{topological}). In Section \ref{gravity-spin-2} we shall see that the gravitational interactions mentioned in the introduction are independent of $T$ and $\overline{T}$ (and obviously of $g'$ because 4D gravitons are not charged under any 6D gauge group).

The GGP solutions solve the 6D supergravity equations in the absence of external sources (other than pure tensions), that is for $I(x)=0$. However, below we will consider perturbations that include couplings between bulk fields and gravitational sources on the brane.

\section{Perturbations and Leading Interaction}

In order to study the massive graviton interactions between sources we have to expand the theory in powers of small fluctuations around the background solutions. In particular, regarding the bulk metric, we substitute 
\be G_{MN} \rightarrow G_{MN} + h_{MN}\label{metricpert}\ee 
 in the action, so that we can interpret $G_{MN}$  as the background metric and $h_{MN}$ as the small fluctuation. We also perform similar replacements for the other bulk fields  (which we do not display for the sake of brevity) and for the brane fields:
\be Y^M\rightarrow Y^M + \xi^M, \, \chi_b\rightarrow \chi_b + \tilde{\chi}_b. \label{branepert}\ee
The mixing (in the linearized theory) between $\tilde{\chi}_b$ and the other perturbations depends on $I(x)$. For the sake of simplicity, we restrict our attention to the cases in which this mixing is absent at the linear level; for example this is true  if we take $I(x)$ to be the standard model Lagrangian with the usual 4D Poincar\'e invariant vacuum.  

The complete bilinear action for the bulk fields and $\xi^M$ has been derived in \cite{Parameswaran:2009bt}. As a consequence of the local symmetries in the initial theory, such bilinear action has the local symmetries
\bea \delta h_{MN} &=& -\eta_{N;M} - \eta_{M;N},  ...  \label{h-gauge} \\
      \delta \xi^M &=&
\eta^M(Y)-\zeta^{\alpha}\partial_{\alpha}Y^{M}, \eea 
where the dots represent the transformation rules of the other bulk fields and $\eta_M=\eta_M(X)$ and $\zeta^\alpha=\zeta^\alpha(x)$ are the gauge parameters associated with the higher dimensional and 4D general coordinate invariance respectively. We use the latter invariance to fix the so called static gauge: $Y^{\mu}= x^{\mu}$ and $\xi^{\mu}=0$. In this gauge we still have the fields $\xi^{m}(x)$, ($ m=4,5$) representing the fluctuations of the brane position along the extra dimensions.

Here we want to include the leading interactions between bulk fields and sources on the branes, which emerge from the action $\int d^4 x \sqrt{-g} I(x)$. To this end we have to perturb the brane metric (\ref{branemetric}) at the linear order in $h_{MN}$ and $\xi^m$:
\bea g_{\mu \nu}& \equiv & G_{MN}(Y(x))\partial_{\alpha}Y^M(x) \partial_{\beta}Y^N(x) \rightarrow G_{\mu \nu}(x,Y_2)+h_{\mu \nu}(x,Y_2)
\nonumber \\ &&+\left(\partial_mG_{\mu \nu}(x,Y_2) \right) \xi^m +2 G_{m\nu}(x,Y_2)\partial_{\mu} \xi^m + ..., \nonumber \\ &=&e^A \eta_{\mu \nu} +h_{\mu \nu}(x,Y_2)+ ...\label{metricexpansion}\eea 
where the dots represent non linear terms and $Y_2\equiv \{Y^m\}$. In the last step in (\ref{metricexpansion}) we used explicitly the fact that\footnote{We also used the fact that the background brane positions $Y_2$ do not depend on $x$, as it is the case for the GGP solutions.} the brane metric at the background level is $e^A \eta_{\mu \nu}$ and $G_{m \mu}=0$ and $\partial_mG_{\mu \nu}(x,Y_2)=0$. The latter property comes from the axisymmetry of the background and from (\ref{Aproperty2}).
Therefore, the brane sources produce the following leading interaction term 
\be \frac{1}{2}\int d^4x \sqrt{-g}\, t^{\mu \nu}(x) h_{\mu \nu}(x, Y_2). \label{brane-bulk}\ee
We call $t^{\mu \nu}$ the brane energy momentum tensor. Notice that we do not include the tension $T$ in the definition of $t^{\mu \nu}$. Notice also that the bulk fields in (\ref{brane-bulk}) are evaluated at the background brane position rather than at the perturbed one $Y^{m}+ \xi^{m}$ and so only bulk fields which have a non vanishing value at background brane positions can contribute to those interactions. 

In order to extract the physical properties of these models one has to fix the bulk local symmetries in addition to the brane ones. Before doing so we have to know the bulk local transformation which leaves (\ref{brane-bulk}) invariant. Such transformation is exactly (\ref{h-gauge}), at least for the GGP solutions (for which (\ref{Aproperty2}) holds) and for covariantly conserved $t^{\mu \nu}$ (that we assume):
$ \nabla_{\mu}t^{\mu \nu} =0$, where $\nabla_{\mu}$ is the covariant derivative computed with the (background) brane metric. Since (\ref{h-gauge}) leaves invariant both the bilinear action and the interaction in (\ref{brane-bulk}), it is that transformation that has to be used in order to fix the bulk gauge symmetries.

\section{Massive Spin-2 Contributions \\ to 4D Effective Gravity}\label{gravity-spin-2}

{\it Leading spin-2 brane interactions}. We want to derive an effective action which describes the exchange of massive gravitons between brane sources. This leads to deviations from Newton's law at short distances. We split (\ref{brane-bulk}) as follows 
\be \frac{1}{2}\int d^4x \sqrt{-g}\,\left[ t^{\mu \nu}(x) \tilde{h}_{\mu \nu}(x, Y_2) + \frac{1}{4}t^{\mu}_{\,\,\mu}(x) h^{\nu}_{\,\,\nu}(x, Y_2)\right],\label{decomposition}\ee
where $h^{\mu}_{\,\,\mu}\equiv e^{-A} \eta^{\mu \nu}h_{\mu \nu}$ and $\tilde{h}_{\mu \nu}\equiv h_{\mu \nu} - e^A \eta_{\mu \nu} h^{\tau}_{\,\, \tau}/4 $ are respectively the trace and the traceless part of $h_{\mu \nu}$ (the indices are raised and lowered by the background metric). We use the bulk local symmetries in (\ref{h-gauge}) to set
\be \partial_{\mu}\tilde{h}^{\mu \nu}=0. \label{gaugeSpin-2}\ee

The advantage of this gauge is that it produces no mixing between $\tilde{h}_{\mu \nu}$ and the other fields, as it can be proved by using the general bilinear Lagrangian in \cite{Parameswaran:2009bt}. An explicit computation gives the following bilinear action for $\tilde{h}_{\mu \nu}$
\be -\frac{1}{4\kappa^2} \int  d^6X \sqrt{-G}\partial_M \tilde{h}_{\mu}^{\,\, \nu} \partial^M \tilde{h}_{\nu}^{\,\, \mu} \label{bilineaSpin-2}.\ee
The field $\tilde{h}_{\mu \nu}$ leads, after dimensional reduction, to the massless graviton and the tower of massive gravitons. Indeed, if we go to the momentum space $\tilde{h}^{\mu \nu}(x)\rightarrow \tilde{h}^{\mu \nu}(p) e^{ipx}$, we can easily see that  a generic mode $\tilde{h}^{\mu \nu}(p)$ has five degrees of freedom (as a consequence of (\ref{gaugeSpin-2}) and the traceless condition) and that these five degrees of freedom can be reduced on-shell\footnote{Here ``on shell'' means that the EOMs in the absence of $t^{\mu \nu}$ have been used.} to two in the massless case by using a residual 4D gauge invariance compatible with (\ref{gaugeSpin-2}). Also, as we discuss below, the 4D mass spectrum emerging from $\tilde{h}_{\mu \nu}(x)$ is exactly that of the spin-2 particles derived in \cite{Parameswaran:2009bt}. The effective action describing the leading spin-2 brane-bulk interactions reads
\be S_{eff}= -\frac{1}{4\kappa^2} \int d^6X \sqrt{-G}\partial_M \tilde{h}_{\mu}^{\,\, \nu} \partial^M \tilde{h}_{\nu}^{\,\, \mu}+\frac{1}{2}\int d^4x \sqrt{-g}\, t^{\mu \nu}(x) \tilde{h}_{\mu \nu}(x, Y_2). \label{effective action}\ee
We should notice that, according to the decomposition (\ref{decomposition}), $h^{\mu}_{\,\,\mu}(x, Y_2)$ also couples to the brane sources (unless the trace of $t^{\mu \nu}$ vanishes). This represents a coupling between bulk scalars and brane sources. The treatment of such interactions requires a way to remove singularities due to codimension two branes, which appear in the scalar sector \cite{Parameswaran:2009bt}.  Here we therefore  consider the graviton-brane couplings only, which will be sufficient to obtain our results.

In order to compute corrections to ordinary 4D gravity we perform a KK expansion for the bulk fields:
\be \tilde{h}_{\mu}^{\,\, \nu}(X)= \sum_{{\bf n},{\bf m}} \tilde{h}_{\mu\,{\bf n,
    m}}^{\,\, \nu}(x)F_{{\bf n,  m}}(u,\varphi),\label{spin-2expansion}\ee
where ${\bf m}$ and ${\bf n}$ are KK
numbers. Since the internal space has the sphere topology we can take $F_{{\bf n,  m}}(u,\varphi)=f_{{\bf n,  m}}(u) e^{i {\bf m }\varphi}$. Also we observe that the bilinear action in (\ref{bilineaSpin-2}) is formally identical to the helicity-2 action in the light-cone gauge analyzed in Ref. \cite{Parameswaran:2009bt}, as expected from 4D Lorentz invariance. So the 4D mass spectrum coming from $\tilde{h}_{\mu \nu}$ is exactly that of the spin-2 particles derived there and we can use the same KK expansion as in Ref. \cite{Parameswaran:2009bt} (for more details see \cite{Parameswaran:2006db}). We summarize here the resulting $f_{{\bf n,  m}}$ and the mass spectrum. 
Suppressing ${\bf n}$ and ${\bf m}$, the explicit
expression for $f\equiv (2N/r_0) e^{-(3A+B)/4} \psi$ is given by
\be \psi =  z^{\epsilon}(1-z)^{\beta}F(a,b,c,z), \label{psi1} 
\ee
where $N$ is a normalization factor, $z\equiv\cos^2\left(u/r_0\right)$, $F$ is Gauss' hypergeometric
function and 
\bea \epsilon &\equiv& \frac{1}{4}\left(1+2|{\bf
    m}|\overline{\omega}\right), \,\, 
\beta\equiv \frac{1}{4}\left(1+2{\bf m}\omega\right),
\,\, c\equiv 1+|{\bf m}|\overline{\omega}, \nonumber \\
a&\equiv&\frac{1}{2}+\frac{{\bf m}}{2}\omega+\frac{|{\bf
    m}|}{2}\overline{\omega} 
+\frac{1}{2}\sqrt{r_0^2M^2+1+{\bf m}^2\left(\omega-\overline{\omega}\right)^2},
\nonumber\\
b &\equiv&\frac{1}{2}+\frac{{\bf m}}{2}\omega+\frac{|{\bf
    m}|}{2}\overline{\omega} 
-\frac{1}{2}\sqrt{r_0^2M^2+1+{\bf m}^2\left(\omega-\overline{\omega}\right)^2},
\label{gbetaabcV}\eea
with
\be \omega\equiv(1-\delta/2\pi)^{-1}, \qquad
\overline{\omega}\equiv(1-\overline{\delta}/2\pi)^{-1}\ee
and
\be M^2= \frac{4}{r_0^2} \left[{\bf n}({\bf n}+1) + \left(\frac{1}{2}
    + {\bf n}\right)|{\bf m}| \left(\omega +\overline{\omega}\right)
  +{\bf m}^2 \omega \overline{\omega}\right] \geq 0,
\label{spin-2-masses}\ee  
where ${\bf n}=0,1,2,3,...\,$. Notice that the functions $\psi$ and the masses $M$ depend on the brane tensions in the non-axisymmetric case (${\bf m}\neq 0$) only.

By inserting (\ref{spin-2expansion})  into $S_{eff}$ %$\tilde{h}_{\mu}^{\,\, \nu}\rightarrow \sqrt{2}\kappa \tilde{h}_{\mu}^{\,\, \nu}$ 
we obtain 
\bea S_{eff} = \int d^4x \sum_{{\bf n,m}}\left\{\frac{1}{4\kappa^2}\left[\tilde{h}_{\mu\,{\bf n,
    m}}^{\,\, \nu}(x)\right]^*(\partial^2 - M_{{\bf n, m}}^2)\tilde{h}_{\nu\,{\bf n,
    m}}^{\,\, \mu}(x) \right.\nonumber \\ \left.+ \frac{1}{\sqrt2 \kappa}\lambda_{{\bf n,m}} \eta_{\nu \tau}t^{\mu \tau}(x) \tilde{h}_{\mu\,{\bf n,
    m}}^{\,\, \nu}(x)\right\}, \label{SeffKK}\eea
where $\partial^2\equiv \eta^{\mu \nu}\partial_{\mu}\partial_{\nu}$ and we chose the following normalization
\be 2\pi \int du  \sqrt{-G} e^{-A}f_{{\bf n,m}}^* f_{{\bf n,m}}=1. \label{normalization}\ee
From the first term in (\ref{SeffKK}) we can see that $M_{{\bf n,m}}$ are the masses of the spin-2 particles, whereas the second term of (\ref{SeffKK}) represents the interaction between the gravitons and the brane sources, whose strength is measured by
\be \lambda_{{\bf n,m}}=\frac{1}{\sqrt{2}} \kappa e^{3 A_b} F_{{\bf n,m \, }b} , \label{lambdanm}\ee
where the label $b$ indicates that the corresponding function is computed at the brane position.

\begin{figure}
\centering
\begin{tabular}{cc}
\includegraphics[scale=0.50]{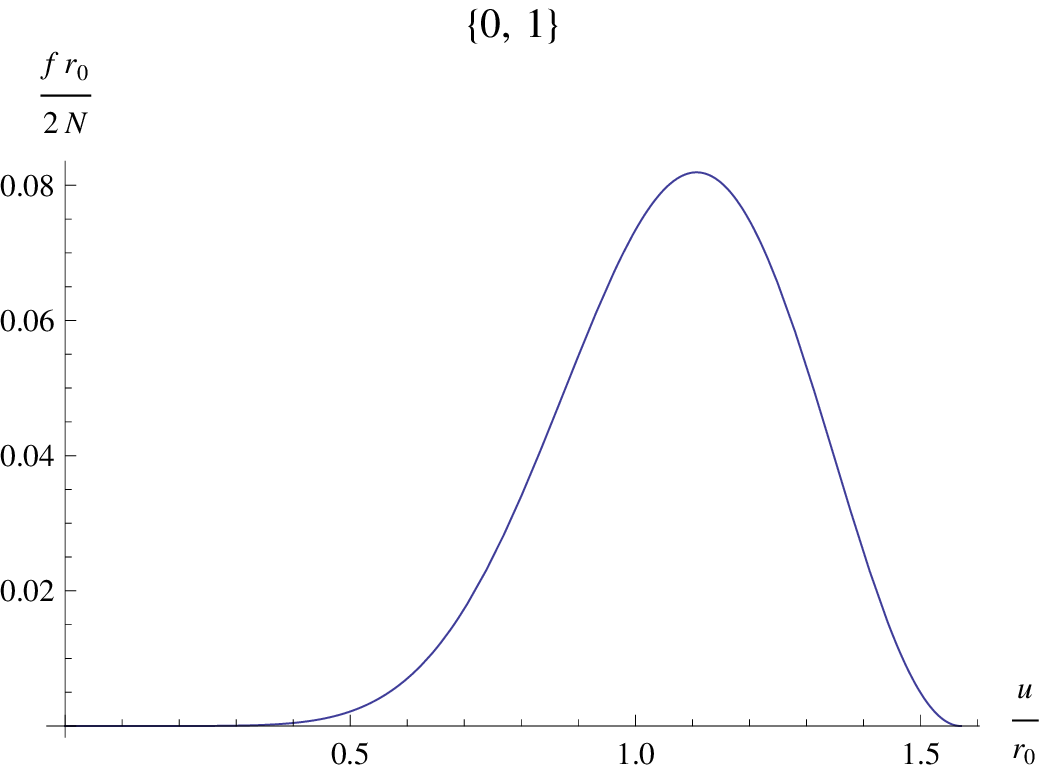} &
\includegraphics[scale=0.50]{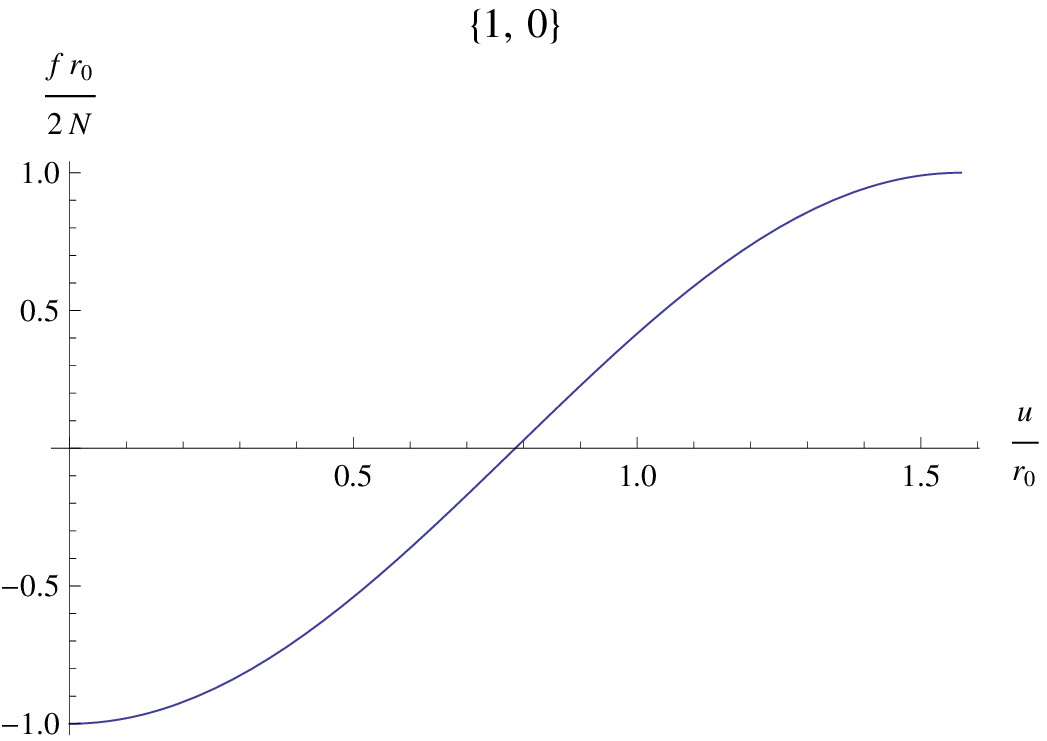}  \\ \\ \includegraphics[scale=0.50]{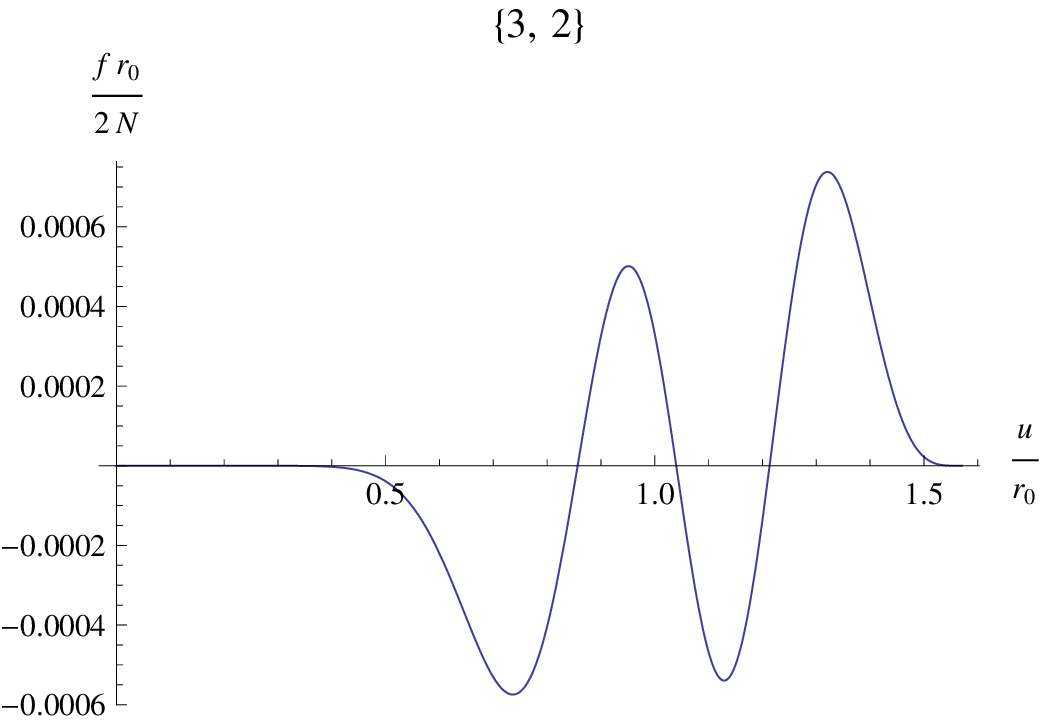} & \includegraphics[scale=0.50]{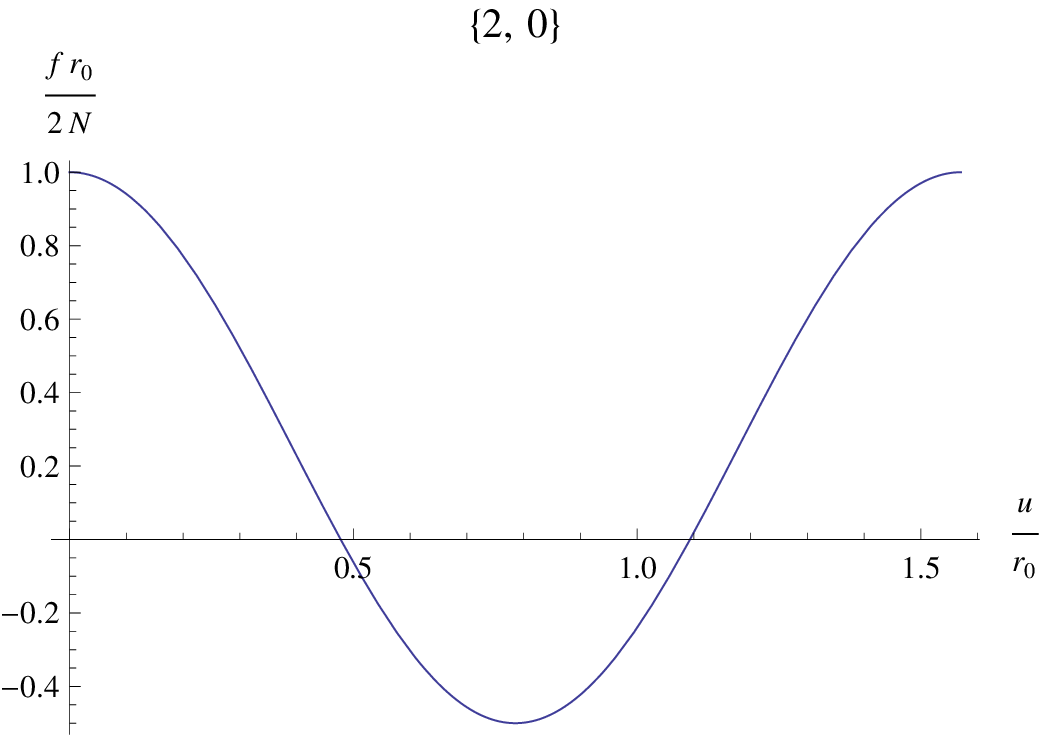} 
\end{tabular}
\caption{\footnotesize Wave functions $f$ (multiplied by $r_0/(2N)$) for different values of $\{ {\bf n,m} \}$. For definiteness we set $\omega=8$ and $\overline{\omega}=2$, which corresponds to deficit angles $\delta=7\pi /4$ and $\overline{\delta} = \pi$. On the right we give two axially symmetric modes, whereas, on the left, two modes that break the axial symmetry. As one can see only the axially symmetric modes are non vanishing on the branes, which are located at $u/r_0=0$ and $u/r_0= \pi /2$. The number of intersections with the $u/r_0$-axis equals ${\bf n}$, in agreement with quantum mechanics.}
\label{fig:wavefns}
\end{figure}

The values of $f_{{\bf n,m \, }b}$ determine which KK modes interact with the brane sources: the effective coupling is indeed $\lambda_{{\bf n,m}}$ and $e^{3A_b}$ is a finite constant (see (\ref{Aproperty1})). An explicit calculation, which makes use of the expression for $f_{{\bf n,m \, }b}$ given above, shows that  $f_{{\bf n,m \, }b}=0$ unless ${\bf m}=0$. This result can be understood remembering  that the  internal space has the $S^2$ topology. A field should be a single valued function at any space-time point, including the north and south poles. In order for this to be true, functions $f$ which depend non trivially on $\varphi$, that is ${\bf m}\neq0$, should go to zero on the poles (see Figure \ref{fig:wavefns}).

Also, a closed form for the couplings $\lambda_{{\bf n,m}}$ can be derived. Let us first consider the brane at $u=\overline{u}$, that is $z=0$. Eq. (\ref{lambdanm}) tells us that we only need the warp factor $e^{A_b}$ evaluated on the brane (which at $u=\overline{u}$ is simply $1$) and the value $f_{{\bf n,m}\,b}$ of the wave functions on the brane:
\be f_{{\bf n,m}\, b}\stackrel{z\rightarrow 0}{\rightarrow}\frac{2}{r_0} \delta_{{\bf m},0} N_{{\bf n, m}=0}e^{-B_b/4} z^{\varepsilon},  \ee
where we used $F(a,b,c,z)\stackrel{z\rightarrow 0}{\rightarrow}1$ and the fact that only axially symmetric modes have non trivial values on the branes and we  showed explicitly the $\{ {\bf n,m} \}$-dependence of the normalization constants. The factor $e^{-B_b/4} z^{\varepsilon}$ is finite and a simple calculation leads to 
\be \lambda_{{\bf n,m}}=\delta_{{\bf m},0} \sqrt{\pi} \kappa_4 N_{{\bf n,m}=0}. \label{step}\ee 
We can also compute explicitly the normalization constants for axially symmetric modes. Condition (\ref{normalization}) can be rewritten as 
\bea && N_{{\bf n,m}=0}= \frac{1}{\sqrt{2\pi I_{{\bf n}}}},\nonumber \\  && I_{{\bf n}}\equiv \frac{2}{r_0}\int_0^{\overline{u}} du \cos\left(u/r_0\right) \sin\left(u/r_0\right)F^2({\bf n}+1,-{\bf n},1,\cos^2\left(u/r_0\right)) 
\eea
and, by using the known formulae $F({\bf n}+1,-{\bf n},1,z)=P_{\bf n}(1-2z)$, where the $P_{\bf n}$ are the Legendre polynomials, and $\int_{-1}^1dyP^2_{\bf n}(y)=2/(2{\bf n}+1)$, we obtain
\be N_{{\bf n,m}=0}=\sqrt{\frac{2{\bf n}+1}{2\pi}}.\ee
Inserting this result into (\ref{step}) leads to a closed form for the couplings $\lambda_{{\bf n,m}}$. A similar procedure can be used to determine the couplings for the brane at $u=0$, the only difference is that there we have a non trivial value of the warp factor $e^{A_b}=\sqrt{\omega/\overline{\omega}}$. We give the explicit expressions\footnote{In deriving Eq. (\ref{lambda2}) we used $F({\bf n}+1,-{\bf n},1,1)=(-1)^{\bf n}$.}:
\bea \lambda_{{\bf n,m}}&=& \sqrt{\frac{2{\bf n} + 1}{2}} \delta_{{\bf m},0} \,\kappa_4, \quad \mbox{at}
\,\, u=\overline{u} \label{lambda1}\\
\lambda_{{\bf n,m}}&=& (-1)^{{\bf n}}\sqrt{\frac{2{\bf n} + 1}{2}}  \delta_{{\bf m},0}\left(\frac{\omega}{\overline{\omega}}\right)^{3/2}\kappa_4, \quad  \mbox{at}
\,\, u=0\label{lambda2}\eea
We observe that the effective 4D gravitational coupling at the $u=0$ brane is rescaled with respect to that at $u=\overline{u}$ by a factor which depends on $\omega/\overline{\omega}$. This is the usual redshift/blueshift due to the warp factor that in our case is trivial at $u=\overline{u}$ and non-trivial at $u=0$.

\vspace{0.3cm}{\it Independence of the brane tensions.}
We now observe an interesting phenomenon in this class of compactifications. Let us first consider the $u=\overline{u}$ brane. The leading spin-2 interactions between brane sources turn out to be independent of the brane vacuum energies $T$ and $\overline{T}$ (among the parameters in (\ref{parametersAC})), because the sole contribution to such interactions comes from the axisymmetric modes, which have both\footnote{The couplings $\lambda_{{\bf n,m}} $ and the masses $M_{{\bf n,m}}$ are indeed the only physical parameters appearing in the effective action (\ref{SeffKK}) (to see it rescale $\tilde{h}_{\mu}^{\,\, \nu}\rightarrow \sqrt{2}\kappa \tilde{h}_{\mu}^{\,\, \nu}$).} $\lambda_{{\bf n,m}}$ and $M_{{\bf n,m}}$ independent of $\omega$ and $\overline{\omega}$. Such property also implies that these gravitational interactions have exactly the same form as in the round sphere case, which coincides with the GGP solutions only in the limit $\omega\rightarrow 1$ and $\overline{\omega}\rightarrow 1$.

 A physical consequence of this is that the brane vacuum energy decouple from such interactions, and the latter turn out to be the same as in the probe brane limit. All the quantities that can be derived from the interactions we have studied (including modifications to Newton's law, which we discuss below) will turn out to be independent of the brane vacuum energy, and therefore no tuning of the tensions can be produced by requiring these quantities to match experiments and observations. 

Finally we observe that the same result remains valid for the $u=0$ brane up to the redshift/blueshift of the 4D gravitational coupling, which disappears in the unwarped limit.  

\vspace{0.3cm}

{\it Deviations from Newton's law.}
We are now interested in the potential generated by a slowly moving particle with mass $M$ on a brane, in the limit of weak stationary fields, that is the Newton's law (and its corrections due to KK fields). To compute this we should look at the fluctuation $H_{\mu \nu}(x,Y_2)$ of the Minkowski metric $\eta_{\mu \nu}$ on the brane. From (\ref{metricpert}) and the metric in (\ref{GGPsolution}) we see that $H_{\mu \nu}\equiv e^{-A} h_{\mu \nu}$. Since $h_{\mu \nu}$ is a bulk field, so is $H_{\mu \nu}$ and it  can be expanded as a sum over the KK fields. However, here we are only interested in the stationary field $H_{00}(\vec{x},Y_2)$ because it is related to the gravitational potential $V$ through the well-known relation
\be V(\vec{x})=-\frac{1}{2}H_{00}(\vec{x},Y_2),\label{standard}\ee
The KK expansion of the bulk field $H_{00}(x,u,\varphi)$ is 
\be H_{00}(x,u,\varphi)= -\sum_{{\bf n,m}}h_{0\,{\bf n,
    m}}^{\,\, 0}(x)f_{{\bf n,  m}}(u) e^{i{\bf m}\varphi} \label{00decomposition}\ee
The fact that we have a slowly moving particle on a brane means that we can approximate the brane energy momentum tensor with $t^{\mu \nu}(\vec{x})=\delta^{\mu}_0 \delta^{\nu}_0 m \,\delta^{(3)}(\vec{x})$, where $m$ is the mass of the particle. A source generates graviton fields according to the equation
\be (\partial^2-M^2_{{\bf n,  m}})\tilde{h}_{\nu \,{\bf n, m}}^{\,\, \mu} = -\sqrt2 \kappa \lambda_{{\bf n,  m}} \eta_{\nu \tau} \tilde{t}^{\mu \tau}, \ee
which can be derived from the action in (\ref{SeffKK}) by means of the minimal action principle. Also $\tilde{t}^{\mu \nu}\equiv t^{\mu \nu}-\eta^{\mu \nu}\eta_{\lambda \sigma} t^{\lambda \sigma}/4$. In particular the $00$-component of the graviton equation with the source $t^{\mu \nu}(\vec{x})=\delta^{\mu}_0 \delta^{\nu}_0 m\, \delta^{(3)}(\vec{x})$  generates stationary KK fields satisfying
\be  (\vec{\nabla}^2-M^2_{{\bf n,  m}})\tilde{h}_{0 \,{\bf n, m}}^{\,\, 0} (\vec{x})= \frac{3}{4} \sqrt2 \kappa \lambda_{{\bf n,  m}}m\, \delta^{(3)}(\vec{x}), \ee
where $\vec{\nabla}^2$ is the (3D) spacial Laplacian on flat space.  Therefore, by using a standard technique,
\be \tilde{h}_{0 \,{\bf n, m}}^{\,\, 0} (\vec{x})= -\frac{3\sqrt2}{16 \pi} \kappa \lambda_{{\bf n,m}} m \frac{e^{-M_{{\bf n,m}}r}}{r}, \ee 
where $r\equiv |\vec{x}|$.
If we now use Eqs. (\ref{standard}) and (\ref{00decomposition}), and we remember that only axially symmetric wave functions are non vanishing on the branes, we deduce the following deviation $\Delta V$ of the Newton's potential due to massive gravitons
\be \Delta V(r) = -\frac{1}{8\pi} \frac{m}{r}\sum_{{\bf n=1}}^{\infty} \frac{3}{2}e^{-3A_b}\lambda^2_{{\bf n, m}=0}e^{-M_{{\bf n,m}=0}r}. \ee
The latter expression is valid  both for the brane at $u=\overline{u}$ and for that at $u=0$ and, by using Eqs. (\ref{lambda1}) and (\ref{lambda2}) we respectively obtain ($G\equiv \kappa_4^2/(16\pi)$)
\bea \Delta V(r) &=& -\frac{Gm}{r}\sum_{{\bf n}=1}^{\infty} \left(3{\bf n} + \frac{3}{2}\right) e^{-2\sqrt{{\bf n}({\bf n}+1)}\,r/r_0}, \quad \mbox{at}
\,\, u=\overline{u}, \\
\Delta V(r) &=& -\frac{Gm}{r}\left(\frac{\omega}{\overline{\omega}}\right)^{3/2}\sum_{{\bf n}=1}^{\infty} \left(3{\bf n} + \frac{3}{2}\right) e^{-2\sqrt{{\bf n}({\bf n}+1)}\, r/r_0},  \quad  \mbox{at}
\,\, u=0.\eea
We see that these corrections to Newton's law confirm the fact that the $T$- and $\overline{T}$-dependence falls out in the final result (up to the redshift/blueshift\footnote{Indeed $\omega/\overline{\omega}=(1-\kappa^2\overline{T}/(4\pi))/(1-\kappa^2T/(4\pi))$.}
 $\left(\omega/\overline{\omega}\right)^{3/2}$ due to the warping) and so no tuning of the brane vacuum energy can be introduced by matching this type of models with the short distance precision tests of gravity, if our 4D phenomena take place on the unwarped brane. Clearly this property remains true if we consider an extended mass distribution rather than a particle.

\section{Final Remarks}

We have computed the leading effective interactions of massive gravitons with 3-brane sources in a 6D brane world scenario which includes the Salam-Sezgin model (and its anomaly free extensions) with its most general solution compatible with 4D Poincar\'e invariance and internal axial symmetry. The internal manifold has a spherical topology and this leads to a discrete KK spectrum with a finite mass gap.

 Up to the usual redshift/blueshift due to the warping (which is trivial on one brane), these interactions turn out to be independent of the brane tensions. In particular this implies that no tuning of the brane tensions are required to match observations and experiments with the above-mentioned interactions. A key feature that has led to this result is the codimension two nature of our branes, and the fact that they live on a compact space (the compactness is required to have a discrete spectrum).  Indeed, the same phenomenon does not arise in the classic codimension one scenarios such as Randall-Sundrum models, where e.g. short distance precision tests of gravity lead to bounds on the brane tensions. These results support the idea that in such models the gravitational interactions are decoupled from the brane vacuum energy. 

Further remarks are now in order.

 If $t_{\,\,\,\mu}^{\mu}\neq 0$, brane interactions mediated by 4D scalars coming from the bulk generically emerge. The proper treatment of this contribution requires a way to remove the singularities due to the codimension two branes, which emerge in the scalar sector \cite{Parameswaran:2009bt}. Different ways of dealing with these singularities can lead to different physical predictions; therefore,  we focused here on the sector (graviton interactions) that is independent of the regularization procedure. There are nevertheless physical situations where one can take $t_{\,\,\,\mu}^{\mu}= 0$, such as the light bending due to  massive sources. 

Also, this type of supergravity models predicts classically massless scalar particles, which could modify the large distance behaviour of gravity. A way to lift these marginal directions, and deal at the same time with the singularities sourced by codimension two branes, has been proposed in \cite{Burgess:2007vi}. However, some difficulties emerge in reconciling this approach with the smallness of the cosmological constant \cite{Arroja:2007ss}. We leave this point as an open issue.

\vspace{0.5cm}

{\bf Acknowledgments.} The author gratefully acknowledges valuable discussions with S. Randjbar-Daemi. This work has been 
supported by  CICYT-FEDER-FPA2008-01430.

\end{document}